\documentclass[prb,twocolumn,showpacs,preprintnumbers,superscriptaddress,amsmath,amssymb]{revtex4}  
\usepackage{graphicx}
\usepackage{dcolumn}

\begin{document}

\title{Heat capacity study of BaFe$_{2}$As$_{2}$: effects of annealing}

\author{C. R. Rotundu}
\email[E-mail address: ]{CRRotundu@lbl.gov}
\affiliation{Materials Sciences Division, Lawrence Berkeley National
Laboratory, Berkeley, CA 94720, USA}
\author{B. Freelon}
\affiliation{Department of Physics, University of California, Berkeley, CA 94720, USA}
\author{T. R. Forrest}
\affiliation{Department of Physics, University of California, Berkeley, CA 94720, USA}
\author{S. D. Wilson}
\affiliation{Department of Physics, Boston College, Chestnut Hill, MA 02467, USA}
\author{P. N. Valdivia}
\affiliation{Department of Materials Science and Engineering, University of California, Berkeley, CA 94720, USA}
\author{G. Pinuellas}
\affiliation{Department of Physics, University of California, Berkeley, CA 94720, USA}
\author{A. Kim}
\affiliation{Department of Physics, University of California, Berkeley, CA 94720, USA}
\author{\\J.-W. Kim}
\affiliation{Advanced Photon Source, Argonne National Laboratory, Argonne, IL 60439, USA}
\author{Z. Islam}
\affiliation{Advanced Photon Source, Argonne National Laboratory, Argonne, IL 60439, USA}
\author{E. Bourret-Courchesne}
\affiliation{Life Sciences Division, Lawrence Berkeley National Laboratory, Berkeley, CA 94720, USA}
\author{N. E. Phillips}
\affiliation{Materials Sciences Division, Lawrence Berkeley National
Laboratory, Berkeley, CA 94720, USA}
\affiliation{Department of Chemistry, University of California, Berkeley, California 94720, USA}
\author{R. J. Birgeneau}
\affiliation{Materials Sciences Division, Lawrence Berkeley National Laboratory, Berkeley, CA 94720, USA}
\affiliation{Department of Physics, University of California, Berkeley, CA 94720, USA}
\affiliation{Department of Materials Science and Engineering, University of California, Berkeley, CA 94720, USA}
\date{\today}

\begin{abstract}

Heat-capacity, X-ray diffraction, and resistivity measurements on a high-quality BaFe$_{2}$As$_{2}$ sample show an evolution of the magneto-structural transition with successive annealing periods. After a 30-day anneal the resistivity in the (ab) plane decreases by more than an order of magnitude, to 12 $\mu\Omega$cm, with a residual resistance ratio $\sim$36; the heat-capacity anomaly at the transition sharpens, to an overall width of less than K, and shifts from 135.4 to 140.2 K. The heat-capacity anomaly in both the as-grown sample and after the 30-day anneal shows a hysteresis of $\sim$0.15 K, and is unchanged in a magnetic field $\mu_{0}$H = 14 T. The X-ray and heat-capacity data combined suggest that there is a first order jump in the structural order parameter. The entropy of the transition is reported.

\end{abstract}

\pacs{74.25.Dw,74.25.Ha,74.70.-b,75.30.Fv,75.25.+z}

\maketitle

\section{Introduction}

In the 122 series of Fe pnictide superconductors, superconductivity is produced by doping a parent compound AFe$_{2}$As$_{2}$ (where A = Ba, Sr, Ca, Eu) with electrons, e.g., by substitution of Co \cite{Sefat, Leithe} or Ni \cite{Li} on the Fe sites or holes, e.g., by substitution of K \cite{Rotter2}, Na \cite{Shirage} or Cs \cite{Sasmal} on the A sites. It was also found that under pressure the 122 parents themselves become superconducting \cite{Torikachvili, Alireza, Miclea}. This led to the finding that even substitutions of isovalent elements, for example, P \cite{Ren} for As or Ru \cite{Schnelle}, Ir \cite{XLWang}, Pd \cite{NNi} or Rh \cite{NNi} for Fe, induce superconductivity.

In this series, the magnetic (spin-density wave) and structural (tetragonal to orthorhombic) transitions in the parent compound are coincident; there is a common magneto-structural transition \cite{Rotter1}. The thermodynamic nature of that transition, first order or second order, remains controversial. In addition, wide ranges of the values of the parameters that characterize the heat capacity have been reported \cite{Rotter1, Wang, Ni}. At least for near optimally doped samples, the doping suppresses the magneto-structural transition, and the superconductivity occurs in the high-temperature tetragonal phase of the parent compound. Nevertheless, the heat capacity of the low-temperature orthorhombic phase has occasionally played a role in analyzing the heat capacity of superconducting samples to separate the lattice and electron contributions, and thus influenced the interpretation of the latter.

In this paper we report heat-capacity measurements on a BaFe$_{2}$As$_{2}$ single crystal, as grown and also after various annealing periods. Some preliminary high resolution X-ray diffraction measurements are also reported. The heat-capacity anomaly at the magneto-structural transition evolves in a complicated way with annealing time. Although the peak is already relatively sharp in the sample as grown, it is substantially sharpened and shifted to a higher temperature after a 30-day anneal. In contrast with that trend, the heat capacity peak is actually broadened after shorter annealing times. The sharp transitions in the sample as grown and after the 30-day anneal support the interpretation of the transition as first order. The behavior after intermediate anneals evinces an extreme sample dependence of its manifestation in the heat capacity.

\section{Experimental Procedure}

The measurements were made on a 10.3-mg fragment of a large single crystal of BaFe$_{2}$As$_{2}$ that had been synthesized by a modified self-flux method \cite{Wang}. The measurements were made on the sample as grown, and after successive anneals at 700$^\circ$ C under a low pressure of Ar gas in a sealed quartz tube, for 1, 4, 8, 14, and 30 days. Except for a measurement of the residual resistivity, which was made on a sample with a well defined geometry, all the measurements reported here were made on the same sample. Another piece of the same large crystal was used for neutron diffraction measurements, which have been reported by Wilson $et$ $al.$ \cite{Wilson} Inductively coupled plasma (ICP) and electron microprobe wavelength-dispersive X-ray spectroscopic (WDS) analysis on the as-grown sample confirmed the 1:2:2 stoichiometry. The ICP and WDS measurements gave, respectively, 1:1.990$\pm$0.028:1.995$\pm$0.145, and 1.008$\pm$0.05:1.994$\pm$0.011:1.998$\pm$0.011. X-ray scattering attests to the high quality of the crystal by rocking curves having full widths at half maxima (FWHM) of about 0.1$^\circ$.
Measurements on the samples were made in Quantum Design apparatus: heat-capacity and resistivity measurements in a Physical Property Measurement System (PPMS), and magnetization in a Magnetic Property Measurement System (MPMS). Synchrotron X-ray diffraction data were collected at the 6-ID beam station of the Advanced Photon Source, Argonne National Laboratory.

\section{Results and Discussion}

Figure 1 shows both the magnetization in a magnetic field of 5 T parallel with the (ab) plane (upper window), and the zero field normalized resistivity in the (ab) plane (lower window) of the as-grown BaFe$_{2}$As$_{2}$ crystal. The very sharp drop of both the magnetization and the resistivity at the magneto-structural transition and the small low temperature magnetization upturn are signs of the high quality of the crystal.

\begin{figure}[h]
\begin{center}\leavevmode
\includegraphics[width=1.1\linewidth]{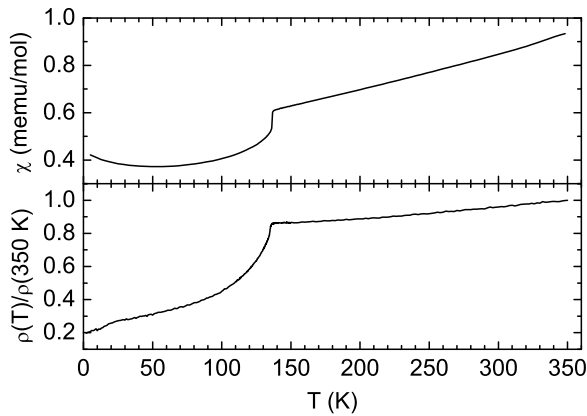}
\caption {Characterization of the as grown BaFe$_{2}$As$_{2}$ single crystal: magnetization in magnetic field of 5 T // (ab) (upper window) and zero field normalized resistivity in (ab) plane (lower window).}\label{fig1}\end{center}\end{figure}

\begin{figure}[h]
\begin{center}\leavevmode
\includegraphics[width=1.1\linewidth]{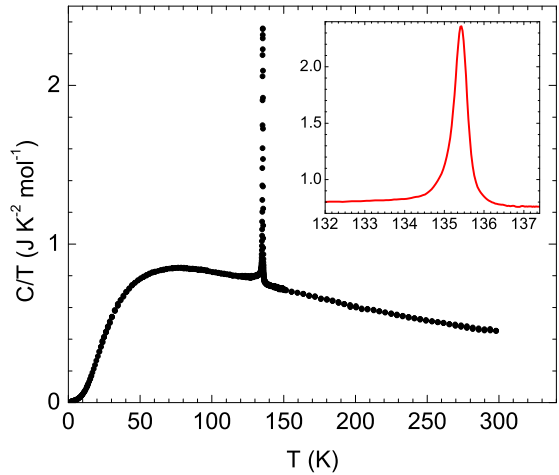}
\caption {The zero-field C/T of the as-grown BaFe$_{2}$As$_{2}$ crystal. The inset shows the same data near the transition on an expanded scale.}\label{fig2}\end{center}\end{figure}
The zero-field heat capacity of the as-grown crystal is shown in Fig. 2. The magneto-structural transition, reflected in the heat capacity by a sharp peak, occurs at 135.4 K, where the heat capacity attains a maximum value of 320 JK$^{-1}$mol$^{-1}$. This value is the highest reported so far on BaFe$_{2}$As$_{2}$ and is another clear indication of the high quality of the sample. The inset shows the same data near the transition on an expanded scale. The ``full width at half maximum'' of the peak is $\sim$0.2 K, well within the window of possible first-order behavior deduced from the neutron scattering measurements of Wilson $et$ $al.$ \cite{Wilson}
\begin{figure}[h]
\begin{center}\leavevmode
\includegraphics[width=1.1\linewidth]{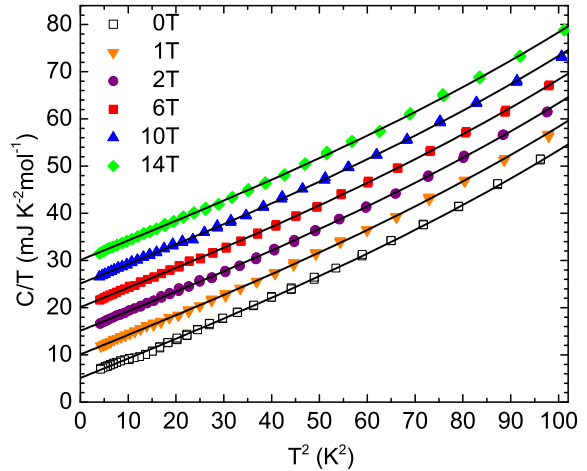}
\caption {The low-temperature specific heat of as grown BaFe$_{2}$As$_{2}$ crystal. The data, in magnetic fields H // c, are displaced by progressive increments of 5 mJK$^{-2}$mol$^{-1}$ for clarity. For each field the curve represents a simultaneous fit to all the data.}\label{fig3}\end{center}\end{figure}

The heat capacities of the as-grown sample below 10 K and in fields up to $\mu_{0}$H = 14 T parallel with the c axis of the crystal are shown in Fig. 3. To the accuracy of the measurements, the heat capacity is independent of field, which shows that the concentration of paramagnetic impurities is well below 10$^{-3}$ mol/mol. The data in all fields were fit simultaneously with the form
\begin{align}
\label{eq: 1}
C&=\gamma T + B_{3}T^{3} + B_{5}T^{5} + B_{7}T^{7}\\
  &= C_{e}+C_{lat}\nonumber,
\end{align}
where the first term on the right represents the electron contribution (C$_{e}$) and other terms are the usual approximation for the lattice contribution (C$_{lat}$). (The deviations of the zero-field points below $\sim$4 K, which are associated with helium adsorbed on the sample and sample platform \cite{QD}, do not make a significant contribution in the overall fit.) The derived values of the coefficients are given in Table 1, where the uncertainty quoted for each parameter is the standard error in the fit. The values of B$_{5}$ and B$_{7}$ have a strong dependence on the temperature interval and number of terms used in the expression for C$_{lat}$, and the uncertainties reflects the interdependence of those terms. However, the reported values do represent the experimental data to within the precision of the PPMS measurements, $\sim$2$\%$. Other reported values of $\gamma$ and B$_{3}$ are included in the table. The experimental data are compared with the resulting fit, where the data in fields other than zero are shifted in C/T by progressive increments of 5 mJK$^{-2}$mol$^{-1}$ for clarity. In each case the common fitting expression is shown, correspondingly shifted.

\begin{table}[h!]
  \begin{center}
\begin{tabular}{|c||c|c|c|c|r|}
	\hline
\text{} & & & & \\
\text{reference} & $\gamma$ & B$_{3}$ & 10$^{5}$B$_{5}$ & 10$^{6}$B$_{7}$\\
\text{} & & & & \\
	\hline \hline
This work   & 5.11$\pm$0.02 & 0.412$\pm$0.003 & 8.3$\pm$11.0 & 6.2$\pm$0.9\\
Ref. 19   & 6.1$\pm$0.3 & 1.51$\pm$0.01 & - & - \\
Ref. 14  & 16(2) & 2.0(5) & - & - \\
Ref. 15  & 37 & 0.60 & - & - \\
	\hline
\end{tabular}
\end{center}
\caption{Coefficients of the low-temperature heat capacity (see Eq. 1). The values of the coefficients are in mJ-K-mol units.}
\end{table}

In order to investigate the nature of the magneto-structural transition, high-resolution X-ray scattering measurements of the tetragonal to orthorhombic structural transition were performed at the Advanced Photon Source's sector 6 beamline. Scattering measurements on the 30-day annealed sample were taken with an X-ray photon energy of 10.452 keV, in a vertical scattering geometry. High resolution reciprocal space scans along the [110] direction across the tetragonal (228)$_T$ reflection, were recorded as the temperature was decreased from 160 K to a base temperature of 6 K. Close to the structural phase transition, measurements were taken at temperature steps of 0.1 K. To ensure that the temperature was stable during each measurement, a rest period of 4 minutes or greater preceded each scan.

\begin{figure}[h]
\begin{center}\leavevmode
\includegraphics[width=1\linewidth,bb=20 210 590 630]{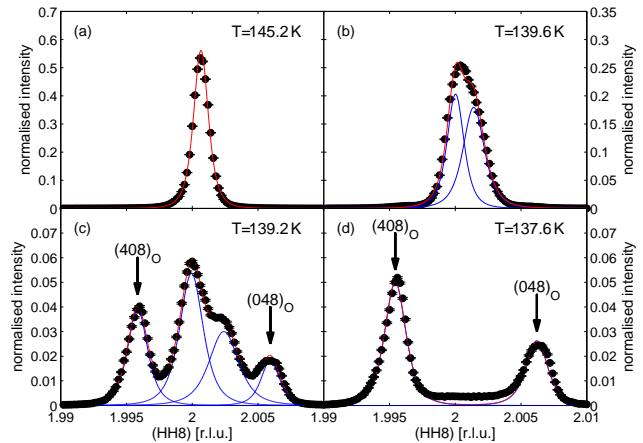}
\caption {Reciprocal lattice scans across the (228)$_T$, (408)$_O$, and (048)$_O$ reflections above and below the structural phase transition. The vertical arrows in panels (c) and (d) show the (408)$_O$ and (048)$_O$ reflections. Panels (b) and (c) show a continuous splitting of the central (228)$_T$ reflection which precedes and coexists with the (408)$_O$ and (048)$_O$ reflections. All reflections are fitted with a Lorentzian squared peak profile.}\label{fig8}\end{center}\end{figure}
Figure 4 shows the evolution of the diffraction profile across the phase transition. Below a temperature of 139.4 K, two features that correspond to the (408)$_O$ and (048)$_O$ orthorhombic reflections appear abruptly. The fact that both orthorhombic reflections are visible, indicates that the sample has orthorhombic domain twinning. These reflections (denoted by vertical arrows in Fig. 4 are independent of the (228)$_T$ reflection, and show only a gradual evolution with decreasing temperature. They are conspicuous at 139.2 K, as shown in Fig. 4(c); they had first appeared, with a much smaller amplitude, at 139.3 K; they were not visible at 139.4 K. This indicates that the structural phase transition is first order in nature. Surprisingly however, this phase transition is preceded, at 140.1 K, by a splitting of the (228)$_T$ reflection, which progresses continuously from zero with decreasing temperature. At 139.6 K the splitting still appears as a broadening of the (228)$_T$ reflection, which can be decomposed into two components, as shown in Fig. 4(b). At 139.2 K it has increased to the point that two reflections are clearly visible. At lower temperatures the amplitude of this reflection decreases rapidly and the splitting continues to increase. This reflection is barely visible at 139.0 K, and not at all at 138.7 K, as shown in Fig. 4(d). It seems probable that the two components approach the orthorhombic reflections, but a quantitative treatment of the kind shown in Figs. 4(b) and 4(c) is not practical. In any case, it is noteworthy that both components of this central reflection disappear as the orthorhombic reflection increase, as expected from a first order transition between two ordered phases. The data in Fig. 4 were all obtained on cooling but the major features are reproduced on warming. This is demonstrated in terms of the structural order parameter in Fig. 5, where both cooling and warming data were included. The very small discontinuities in the order parameters at 141 K (in the region between Figs. 4a and 4b) are not significant in relation to the uncertainty in the decomposition of the central reflection into two components.

Therefore this result indicates that there are in fact two parts to this phase transition. First, below 140 K, the tetragonal structure becomes orthorhombic with a small distortion. The x-ray scattering data shows a splitting of the tetragonal reflection into the two orthorhombic reflections, that starts continuous with temperature. Second, at the slightly lower temperature of 139.4 K, a transition between two orthorhombic structures is observed. This change between the two orthorhombic structures happens abruptly with temperature. Thus, these results are consistent with the thermodynamic measurements that shown a first order transition.
To emphasize this point we present the structural order parameter, $\delta$=(a-b)/(a+b), as a function of temperature in Fig. 5. The lattice constants $a$ and $b$ were determined through fitting the positions of the reflections with a Lorentzian squared profile in a non-linear least squares analysis. Figure 5 shows that the two sets of orthorhombic reflections co-exist for a temperature range of 0.4 K. 
It should also be noted at this point that the ordering temperature of the first-order transition is slightly lower than the temperature indicated by the heat capacity measurements. Possible explanations include slight differences in the calibrations of the respective thermometers and, for the X-ray experiment, a subtle heating of the sample by the X-ray beam which is not recorded by the temperature sensors. It should also be noted that the stability of the temperature during each scan and the reproducibility of these results upon cooling and warming, indicate that a kinetic origin of these effects is unlikely.

\begin{figure}[h]
\begin{center}\leavevmode
\includegraphics[width=1\linewidth,bb=30 205 560 600]{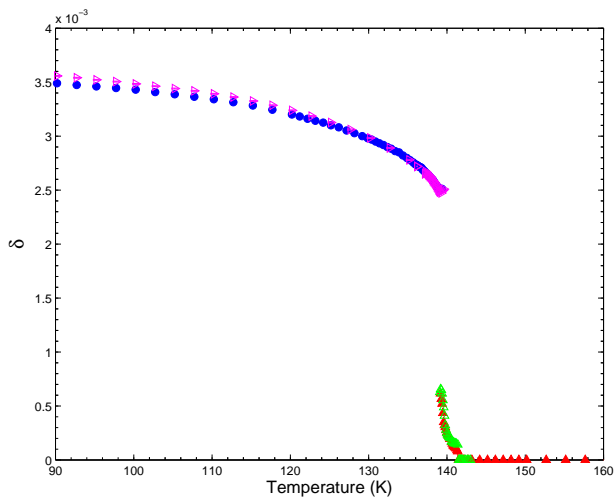}
\caption {The structural order parameter, $\delta$=(a-b)/(a+b), as a function of temperature. Filled symbols are for cooling data and open symbols are for the warming data.}\label{fig8}\end{center}\end{figure}

The unusual behavior of this structural transition may provide an explanation as to why previous diffraction measurements have characterized it as second order. Clearly, with lower angular and temperature resolutions, one could mistake the initial splitting of the tetragonal reflection and the discontinuous appearance of the orthorhombic reflections as being the same thing. In fact Wilson $et$ $al.$ \cite{Wilson} acknowledge this point by stating that the phase transition could be first order within a temperature interval of 0.5 K. (Their measurements of the structural order parameter, by neutron scattering, were made on another piece of the crystal from which the as-grown sample was obtained. They reported a temperature dependence of $\delta$ similar to that shown in Fig. 5.) Whether the effects described in this paper are a universal feature of this phase transition, or just specific to this annealed sample, remains to be determined. However, these results do make a compelling case for further investigations, especially their relationship to the concurrent magnetic phase transition. Specifically, simultaneous high resolution diffraction measurements of both the magnetic and structural order parameters will be necessary to clarify the precise nature of the phase transition(s).

\begin{figure}[h]
\begin{center}\leavevmode
\includegraphics[width=1.1\linewidth]{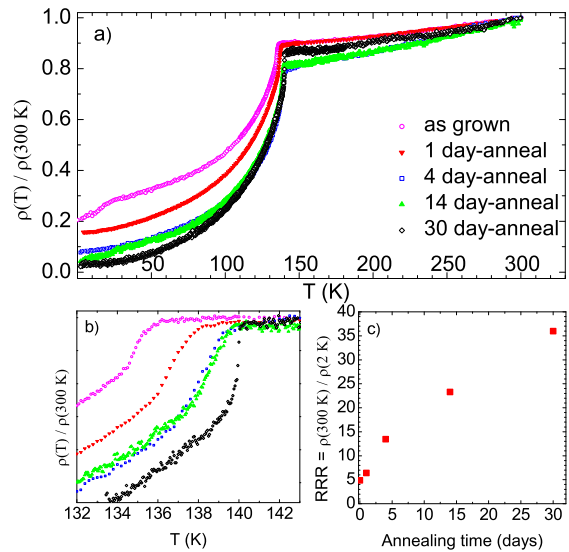}
\caption {(a) Nomalized resistivity of BaFe$_{2}$As$_{2}$ single crystal, as grown and after 1, 4, 14, and 30 days of annealing. (b) The same data as in (a) near the transition. The data were shifted in $\rho$(T)/$\rho$(300 K) to facilitate comparison of the transition temperature after annealing. (c) Residual resistivity ratio of in-plane resistivity, RRR$\equiv$$\rho$(300 K)/$\rho$(2 K), versus annealed time periods.}\label{fig4}\end{center}\end{figure}

Figure 6a shows the temperature dependence of the resistivity in the (ab) plane normalized to the resistivity at 300 K for the as-grown sample and after annealing periods of 1, 4, 8, 14, and 30 days. The transition temperature increases continuously with annealing of up to 4 days, but remains constant after longer annealing periods (Fig. 6b). The resistivity has a weak linear dependence on temperature above the transition but decreases sharply below the transition. This is in disagreement with the $log(1/T)$ behaviour below 12 K reported by Wang $et$ $al$. \cite{Wang}
In Fig. 6c the in-plane residual resistivity ratio, RRR$\equiv$$\rho$(300 K)/$\rho$(2 K), is plotted versus annealing time. While the RRR of the as-grown sample is slightly above 5, one of the highest value reported for the system, it is continuously increased by annealing, attaining a value of 36 after 30 days. The residual resistivity, $\rho$(2 K), of the sample after the 30-day anneal is about 12 $\mu\Omega$cm, one order of magnitude smaller than the lowest value reported so far on the system. This decrease of the residual resistivity can be caused by the healing of the imperfections and dislocations that are centers of scattering through the annealing process.

\begin{figure}[h]
\begin{center}\leavevmode
\includegraphics[width=1.1\linewidth]{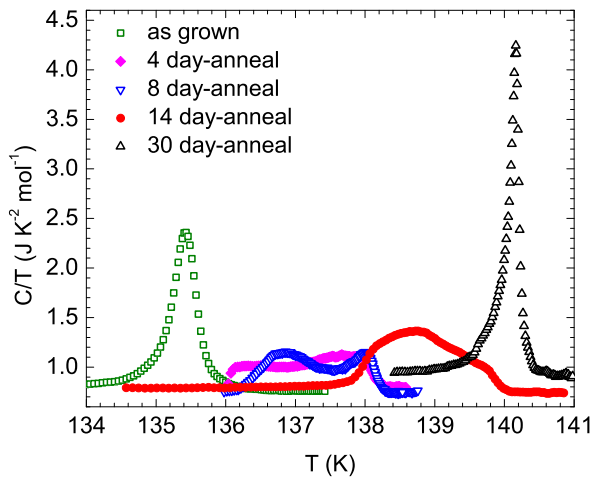}
\caption {C/T near the transitions of the BaFe$_{2}$As$_{2}$ sample, as grown and after annealings.}\label{fig5}\end{center}\end{figure}

The heat capacity near the transition is shown in Fig. 7. After 4 or 8 days of annealing the single sharp peak of the as-grown sample evolves into a broad feature at a somewhat higher temperature with an overall width of $\sim$1.5 K. With a longer annealing time, 14 days, the anomaly moves to still higher temperature, with approximately the same width. After 30 days of annealing the broad feature evolves back into a single sharp peak at 140.2 K, with the heat capacity reaching a maximum of 600 JK$^{-1}$mol$^{-1}$. The entropy change at the transition of the sample, both as grown and after the 30-day anneal is $\sim$0.84 JK$^{-1}$mol$^{-1}$.

\begin{figure}[h]
\begin{center}\leavevmode
\includegraphics[width=1.1\linewidth]{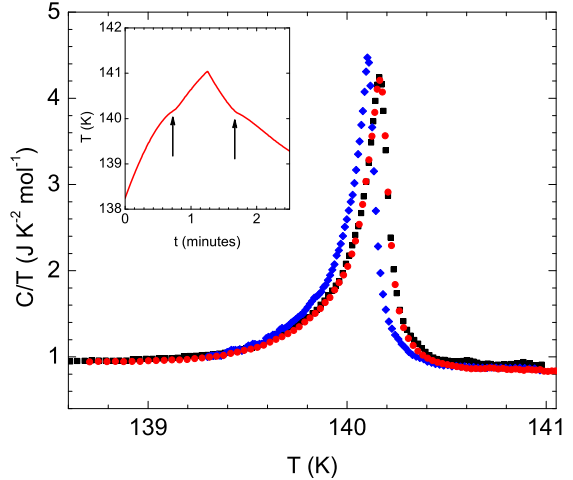}
\caption {C/T near the transitions of the BaFe$_{2}$As$_{2}$ sample after the 30-day anneal in zero field, heating ($\blacksquare$) and cooling ($\blacklozenge$); in 14 T, heating ($\bullet$). The inset shows the relaxation curves of the sample temperature versus time of a measurement spanning the transition. The arrows point to regions of the curve that have a reduced dT/dt, showing the presence of the latent heat, a signature of the first order transition.}\label{fig6}\end{center}\end{figure}

The evolution of the heat-capacity anomaly at the transition with annealing time is unusual. Annealing a sample, at least under suitable conditions, is usually expected to produce a steady improvement in sample quality. The resistivity results are consistent with that expectation. The behavior of the heat-capacity anomaly - broadening at intermediate annealing times, followed by sharpening at longer times - has no obvious explanation. It does, however emphasize the sensitivity of the transition to details of the sample preparation. It is also consistent with measurements on other samples that have shown a variety of heat-capacity anomalies associated with the transition. The reported widths of the specific-heat anomalies, at their ``base'' varies from 1 to 5 K. A variety of peak heights, in C/T measured from the base, and transition temperatures have been reported: ~0.3 JK$^{-2}$mol$^{-1}$ at 138 K \cite{Rotter1}, ~0.9 JK$^{-2}$mol$^{-1}$ at 136 K \cite{Dong}, ~0.1.5 JK$^{-2}$mol$^{-1}$ at 132 K \cite{Budko}, ~0.4 JK$^{-2}$mol$^{-1}$ at 133 K \cite{Sefat2}, and ~0.5 JK$^{-2}$mol$^{-1}$ at 133 K \cite{Fisher}. In several cases the anomalies are comparable to those reported here for intermediate annealing times. In one case \cite{Mandrus} a double peak was reported, i.e., similar in that respect to those reported here for intermediate annealing times.

Figure 8 shows C/T in the vicinity of the transition after 30 days of annealing.
These data (and the corresponding data in Fig. 2) were obtained by an analysis of the heating and cooling segments of individual heating cycles, with a program that was provided with the PPMS for use in the vicinity of sharp features in the heat capacity. The zero-field data derived from the heating and cooling segments show a hysteresis of 0.15 K. The existence of hysteresis is consistent with the conclusion, based primarily on other results, that the transition is first order. (This subtle hysteresis of 0.15 K would not have been detected in the neutron study of Wilson $et$ $al.$ \cite{Wilson}) The inset of Fig. 8 shows the heating and relaxation curves for a heat capacity point that spanned the transition. The arrows point to regions of reduced dT/dt, which show the presence of a latent heat, another indication of a first-order transition \cite{Lashley}. The in-field data show that the transition is unaffected by a magnetic field $\mu_{0}$H = 14 T. This is a surprising result given the combined magnetic and structural nature of the phase transition. The heat capacity of the as-grown sample in the vicinity of its transition showed the same evidence of hysteresis, a latent heat, and the absence of an effect of a magnetic field.

\begin{figure}[h]
\begin{center}\leavevmode
\includegraphics[width=1.1\linewidth]{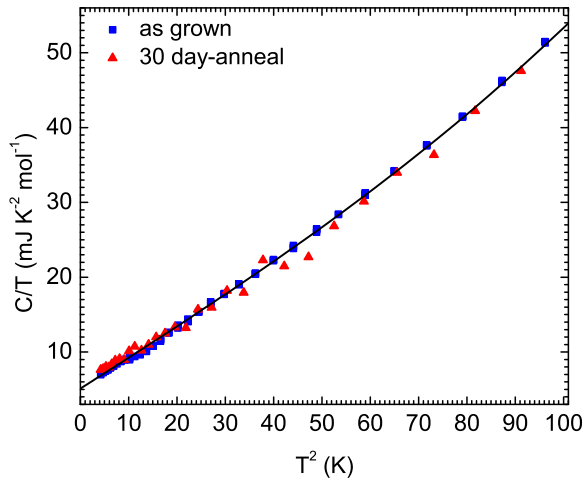}
\caption {Zero-field, low-temperature C/T versus T$^2$ of the as-grown and 30-day-annealed BaFe$_{2}$As$_{2}$ samples. The solid curve represents the collective fit to the data for the as-grown sample in all fields.}\label{fig7}\end{center}\end{figure}

The zero-field low-temperature heat capacity after the 30-day anneal is compared with that of the as-grown sample in Fig. 9. The measurements on the annealed sample were made on $\sim$1/2 of the sample, which accounts for the scatter in the data. To within the precision of the measurements, the parameters that characterize the low temperature heat capacity (see Table 1) are unaffected by the anneal.

\section{Summary}

We made heat-capacity, X-ray-diffraction, and resistivity measurements on a high quality BaFe$_{2}$As$_{2}$ sample as grown and after annealing for 1, 4, 8, 14 and 30 days at 700$^\circ$ C. We observed an evolution of the magneto-structural transition with successive annealing periods. After 30 days of annealing  the resistivity in the (ab) plane decreased to 12 $\mu\Omega$cm, with a residual resistance ratio of $\sim$36; the heat-capacity anomaly at the transition sharpened, to an overall width of less than 1 K, and shifted from 135.4 to 140.2 K. The heat-capacity anomaly in both the as-grown sample and after the 30-day anneal shows a hysteresis of 0.15 K, and is unchanged in a magnetic field $\mu_{0}$H = 14 T. The X-ray and heat-capacity data combined suggest that there is a first order jump in the structural order parameter. This jump is also apparent in regions of reduced dT/dt in the heating and relaxation curves taken in the heat-capacity measurements near the transition. The conduction-electron density of states and the low-temperature lattice heat capacity of the orthorhombic phase are unaffected by the annealing.

\begin{acknowledgments}

This work was supported by the Director, Office of Science, Office of Basic Energy Sciences, U.S. Department
of Energy, under Contract No. DE-AC02-05CH11231 and Office of Basic Energy Sciences US DOE DE-AC03-
76SF008.

\end{acknowledgments}

\end{document}